\documentclass[prd,twocolumn,showpacs,superscriptaddress,preprintnumbers,
amsmath,amssymb]{revtex4-1}

\usepackage{graphicx} 
\usepackage{tikz}
\usetikzlibrary{arrows}

\usepackage[colorlinks=true,urlcolor=blue,linkcolor=blue,citecolor=blue]{hyperref}
\usepackage[normalem]{ulem}
\usepackage[capitalize]{cleveref}

\usepackage[export]{adjustbox}
 
\usepackage{comment}
\usepackage{flushend}

\usepackage[T1]{fontenc} 

\usepackage{mathtools}

\usepackage{xcolor}
\usepackage{graphicx}
\usepackage{dcolumn}
\usepackage{bm}
\usepackage{multirow}

\usepackage{epstopdf}

\newcommand{\Beq}{\begin{equation}\begin{aligned}}
\newcommand{\Eeq}{\end{aligned}\end{equation}}

\begin{document}

\preprint{IPMU23-0033, YITP-23-115, KEK-QUP-2023-0022, KEK-TH-2555, KEK-Cosmo-0325}

\title{Universal Gravitational Waves from Interacting and Clustered Solitons} 

\author{Kaloian D. Lozanov}
\email{kaloian.lozanov@ipmu.jp}
\affiliation{Kavli Institute for the Physics and Mathematics of the Universe (WPI), UTIAS
The University of Tokyo, Kashiwa, Chiba 277-8583, Japan.}

\author{Misao Sasaki}
\email{misao.sasaki@ipmu.jp}
\affiliation{Kavli Institute for the Physics and Mathematics of the Universe (WPI), UTIAS
The University of Tokyo, Kashiwa, Chiba 277-8583, Japan.}
\affiliation{Center for Gravitational Physics, Yukawa Institute for Theoretical Physics,
Kyoto University, Kyoto 606-8502, Japan}
\affiliation{Leung Center for Cosmology and Particle Astrophysics, National Taiwan
University, Taipei 10617, Taiwan}

\author{Volodymyr Takhistov}
\email{vtakhist@post.kek.jp}
\affiliation{International Center for Quantum-field Measurement Systems for Studies of the Universe and Particles (QUP, WPI),
High Energy Accelerator Research Organization (KEK), Oho 1-1, Tsukuba, Ibaraki 305-0801, Japan}
\affiliation{Theory Center, Institute of Particle and Nuclear Studies (IPNS), High Energy Accelerator Research Organization (KEK), Tsukuba 305-0801, Japan
}
\affiliation{Graduate University for Advanced Studies (SOKENDAI), \\
1-1 Oho, Tsukuba, Ibaraki 305-0801, Japan}
\affiliation{Kavli Institute for the Physics and Mathematics of the Universe (WPI), UTIAS, \\The University of Tokyo, Kashiwa, Chiba 277-8583, Japan}

\date{\today}

\begin{abstract}
Causal soliton formation (e.g. oscillons, Q-balls) in the primordial Universe is expected to give rise to a universal gravitational wave (GW) background, at frequencies smaller than scales of nonlinearity. We show that modifications of the soliton density field, driven by soliton interactions
or initial conditions, can significantly enhance universal GWs. Gravitational clustering of solitons naturally leads to
generation of correlations in the large-scale soliton density field. As we demonstrate for axion-like particle (ALP) oscillons, the growing power spectrum amplifies universal GW signals, opening new avenues for probing the physics of the early Universe with upcoming GW experiments. Our results are applicable to variety of scenarios, such as solitons interacting through a long range Yukawa-like fifth force.
\end{abstract}

\maketitle

{\it Introduction}.--- The extreme conditions in the early Universe provide an ideal laboratory for the production of solitonic non-perturbative objects beyond particles that can arise in a broad variety of cosmological theories~\cite{Vilenkin:2000jqa,Shnir:2018yzp}. The high temperatures, potentially far exceeding TeV-scales at times before the era of Big Bang Nucleosynthesis (BBN), readily allow for phase transitions in theories going beyond the Standard Model (BSM). This can lead to the copious generation, through the Kibble-Zurek mechanism \cite{Kibble:1976sj,Zurek:1985qw}, of cosmic defects such as monopoles (dimension 0), cosmic strings (dimension 1) or
domain walls (dimension 2). Furthermore, spectator fields that are energetically subdominant during the early stages of cosmic inflation could also form solitonic objects before BBN~\cite{Amin:2014eta}. Even if spectator fields never enter thermal equilibrium with the primordial plasma in the period between the end of inflation and the onset of BBN, their self-interactions can lead to non-thermal phase transitions and the formation of solitonic structures such oscillons~\cite{Amin:2011hj}, with intriguing possible phenomenological implications such as formation of primordial black holes~\cite{Cotner:2018vug,Cotner:2019ykd}.

The soliton formation is always accompanied by generation of a stochastic gravitational wave (GW) background (e.g.~\cite{Lozanov:2019ylm,Caprini:2018mtu}). The spectrum of the GW background is peaked at a characteristic frequency determined by typical non-linear scales (i.e. correlation lengths) at the time of soliton formation. Such generated GW signals establish fruitful avenues for probing physics at high energies relevant to the primordial Universe.

Recently~\cite{Lozanov:2023aez}, it was found that the peaked GW background from soliton formation is expected to be universally accompanied by a GW background at lower frequencies, lying at frequencies orders of magnitude below the characteristic frequencies associated with non-linearity scales. This universal GW (UGW) background is sourced by the Poissonian tail in the power spectrum of the density field of the solitons on length-scales greater than the correlation length. The mechanism is expected to be relevant for the majority of causal soliton formation scenarios, regardless of whether the nature of the solitons is topological or non-topological, or whether the transition is thermal or non-thermal.

In this work we explore how interactions of solitons affect UGW emission. Soliton interactions affect evolution of the Poissonian tail in the density field. In particular, generic clustering of solitons due to gravity can generate correlations on large, but still sub-horizon, scales. We demonstrate that alterations in the spatial correlations of the soliton density field, reflected by a change in its power spectrum, can result in significant modification of the expected UGW signatures.

{\it Universal gravitational waves.}--- Consider a causal formation of solitons in the radiation dominated primordial Universe, in the period between the end of inflation and the beginning of BBN, sourced by a subdominant spectator field $\phi$. 
At much larger length-scales than those associated with solitons, considering that the formation of solitonic objects are statistically independent events, the density field obeys Poissonian statistics.
This implies that associated power spectrum follows behavior $\propto k^3$. As shown in Ref.~\cite{Lozanov:2023aez}, this universal feature in the power spectrum of causally formed solitons sources GW signals at frequencies smaller than the peak frequency due to nonlinear dynamics. This IR extension of the stochastic background has been dubbed ``universal GWs''~\cite{Lozanov:2023aez}. 

We briefly overview the main features of UGWs, following Ref.~\cite{Lozanov:2023aez}. We start with a perturbed flat
Friedmann-Lemaıtre-Robertson-Walker (FLRW) metric~\cite{Kodama:1984ziu} and consider that Universe is dominated by radiation fluid density $\rho_r$ and subdominant matter density $\rho_{\phi}$. Then, through some mechanism like those mentioned above, the subdominant matter component from $\phi$ leads to formation of massive solitons at time $\tau_i$. Solitons that are long-lived come to matter dominate the Universe and then decay. 

During radiation domination, isocurvature perturbations can be described as~\cite{Kodama:1986ud} 
\begin{equation}
   S = \dfrac{\delta \rho_{\phi}}{\rho_{\phi}} - \dfrac{3}{4} \dfrac{\delta \rho_r}{\rho_r}~.
\end{equation}
We take $k$ as the co-moving wavenumber, $x=k\tau$ as time coordinate and $\kappa=k/k_{\rm eq}$ with $k_{\rm eq}$ being the mode that enters the horizon at matter-radiation equality. Then, on subhorizon perturbation scales relevant for solitons (i.e. $x \gg 1$) and inside radiation-domination regime (i.e. $x/\kappa = k_{\rm eq}\tau \ll 1$), curvature perturbations $\Phi$ and isocurvature perturbations can be approximately described as~\cite{Lozanov:2023aez} (for superhorizon initial conditions see Ref.~\cite{Domenech:2021and})
\Beq\label{eq:sphisol}
&\Phi\simeq \frac{3S_{\bf k}(\tau_i)}{2\sqrt{2}\kappa} \frac{1}{x^3}\left[6+x^2-2\sqrt{3}x\sin\left(\frac{x}{\sqrt{3}}\right)-6\cos\left(\frac{x}{\sqrt{3}}\right)\right]\,,\\
&S\simeq S_{\bf k}(\tau_i)+\frac{3S_{\bf k}(\tau_i)}{2\sqrt{2}\kappa}\left[x+\sqrt{3}\sin\left(\frac{x}{\sqrt{3}}\right)-2\sqrt{3}{\rm Si}\left(\frac{x}{\sqrt{3}}\right)\right]\,,
\Eeq
where Si$(x)$ is the sine-integral function and subscript ${\bf k}$ describes the dependence on the initial conditions. 

From Eq.~\eqref{eq:sphisol} it is apparent that isocurvature perturbations directly source curvature perturbations $\Phi$.
During soliton formation, one generically expects $\delta\rho_\phi/\rho_\phi\gg\delta\rho_r/\rho_r$, with energy density power spectrum of $\phi$ being approximately isocurvature $\mathcal{P}_S$. The production of approximately Poissonian isocurvature on large (sub-horizon) scales, can be converted into growing curvature perturbations as the solitons are redshifted more slowly than the background radiation.  

The generated curvature from soliton isocurvature sources induced GWs at second order, giving rise to the universal background~\cite{Lozanov:2023aez}.
We can define the initial dimensionless power spectrum of isocurvature fluctuations $\langle S_{\bf k}(\tau_i)S_{\bf k'}(\tau_i)\rangle=(2\pi^2/k^3)\mathcal{P}_S(k)\delta^{(3)}({\bf k}+{\bf k'})$.  The resulting induced GWs at time $x_c$ from initial isocurvature perturbations are given by \cite{Domenech:2021and,Lozanov:2023aez}
\Beq\label{eq:GaussIso}
\Omega_{\rm GW,c}(k)=\frac{2}{3}\int_0^\infty &dv\int_{|1-v|}^{1+v}du\left[\frac{4v^2-(1-u^2+v^2)^2}{4uv}\right]^2\\
&\times\overline{I^2(x_c,k,u,v)}{\mathcal{P}_{S}(ku)}{\mathcal{P}_{S}(kv)}\,,
\Eeq
where $\overline{I^2}$ is the oscillation average of the square of the kernel. For sub-horizon modes at the time of soliton formation, we need to consider kernel for a finite initial time  $x_i = k \tau_i$~\cite{Lozanov:2023aez}.

The produced GW spectrum at present day is given by 
\Beq \label{eq:gwtoday}
\Omega_{\rm GW,0}(k)= \Omega_{r,0}\left(\frac{g_{*}(T_{c})}{g_{*,0}}\right)\left(\frac{g_{* s}(T_{c})}{g_{* s,0}}\right)^{-4/3}\Omega_{\rm GW,c}(k)\,,
\Eeq
where $g_{*}(T_{c})$, $g_{* s}(T_{c})$ are the effective number of degrees of freedom in the energy and entropy densities at $T_c$, with $g_{*,0}= 3.36$, $g_{* s,0} = 3.91$ being values at present and $\Omega_{r,0}\simeq4.18\times 10^{-5}h^{-2}$ is the fractional density of radiation at present~\cite{Planck:2018vyg}. We consider the factors in Eq.~\eqref{eq:gwtoday} depending on $g_{*,0}$, $g_{* s,0}$ to be $\mathcal{O}(1)$.

{\it Correlations of interacting solitons}.--- 
We now illustrate the effects of soliton interactions on UGWs by extending the analysis of Ref.~\cite{Lozanov:2023aez} to include realistic gravitational effects on the evolution of the soliton density field on large scales expected to be applicable for all types of solitons. We account for the modifications of the soliton power spectrum in the infrared (IR) regime due to soliton long-range interactions and demonstrate the effects of gravitational clustering of solitons on the universal GWs.

We define the soliton correlation function as $\xi(r)=\langle S(r)S(0)\rangle=\int d^3k (2\pi^2/k^3)P_S(k)e^{i{\bf k}\cdot{\bf r}}$. As demonstrated by numerical simulations (e.g.~\cite{Amin:2019ums}), gravitational clustering of solitons, such as oscillons, modifies the large scale correlations from Poissonian behavior $\xi(r)\propto \delta(r)$ to $\xi(r)\propto r^{-2}$. 

\begin{figure*}[t]
\includegraphics[trim={0 0 0cm 0cm},clip, width=0.48\textwidth,valign=c]{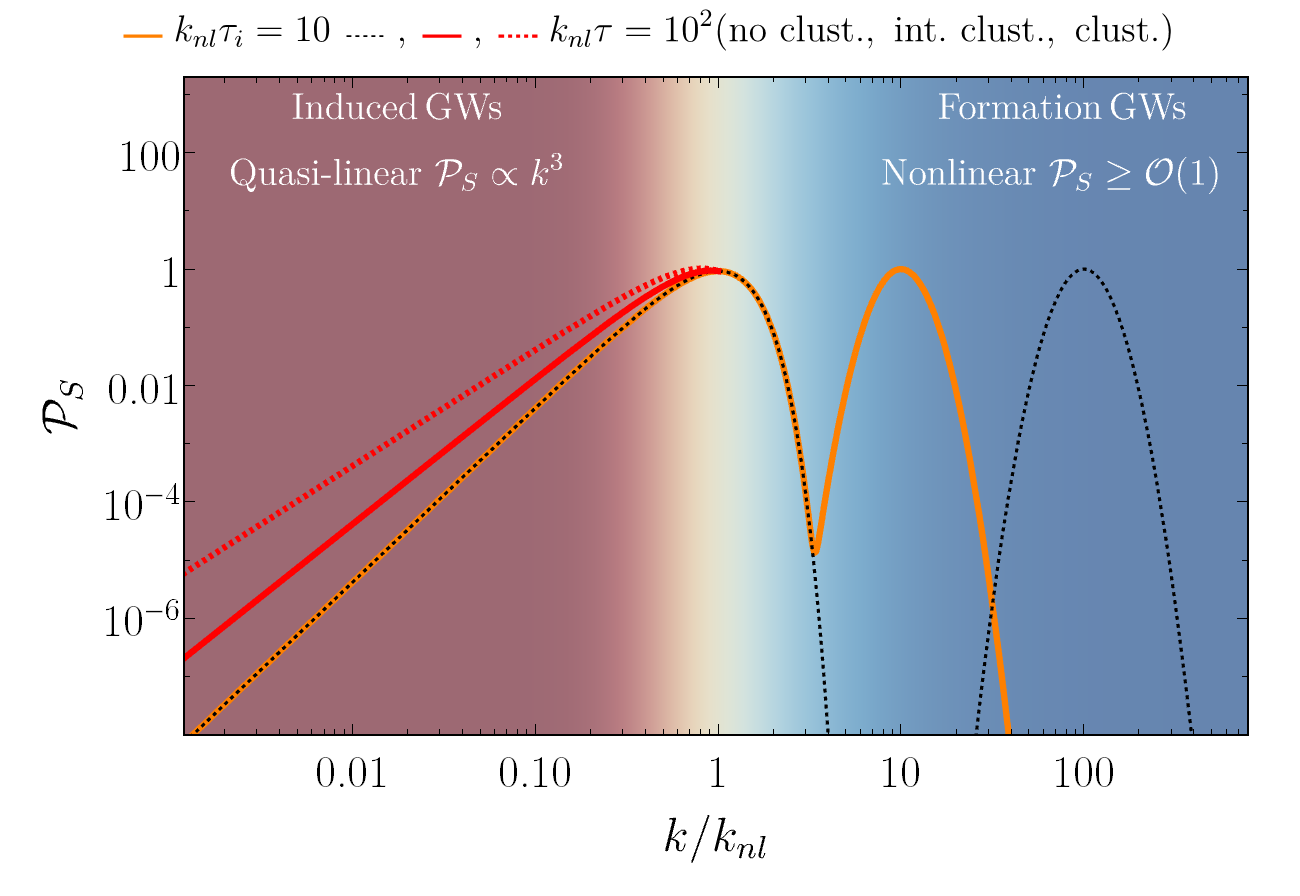}
\includegraphics[trim={0 0 0cm 0.4cm},clip, width=0.49\textwidth,valign=c]{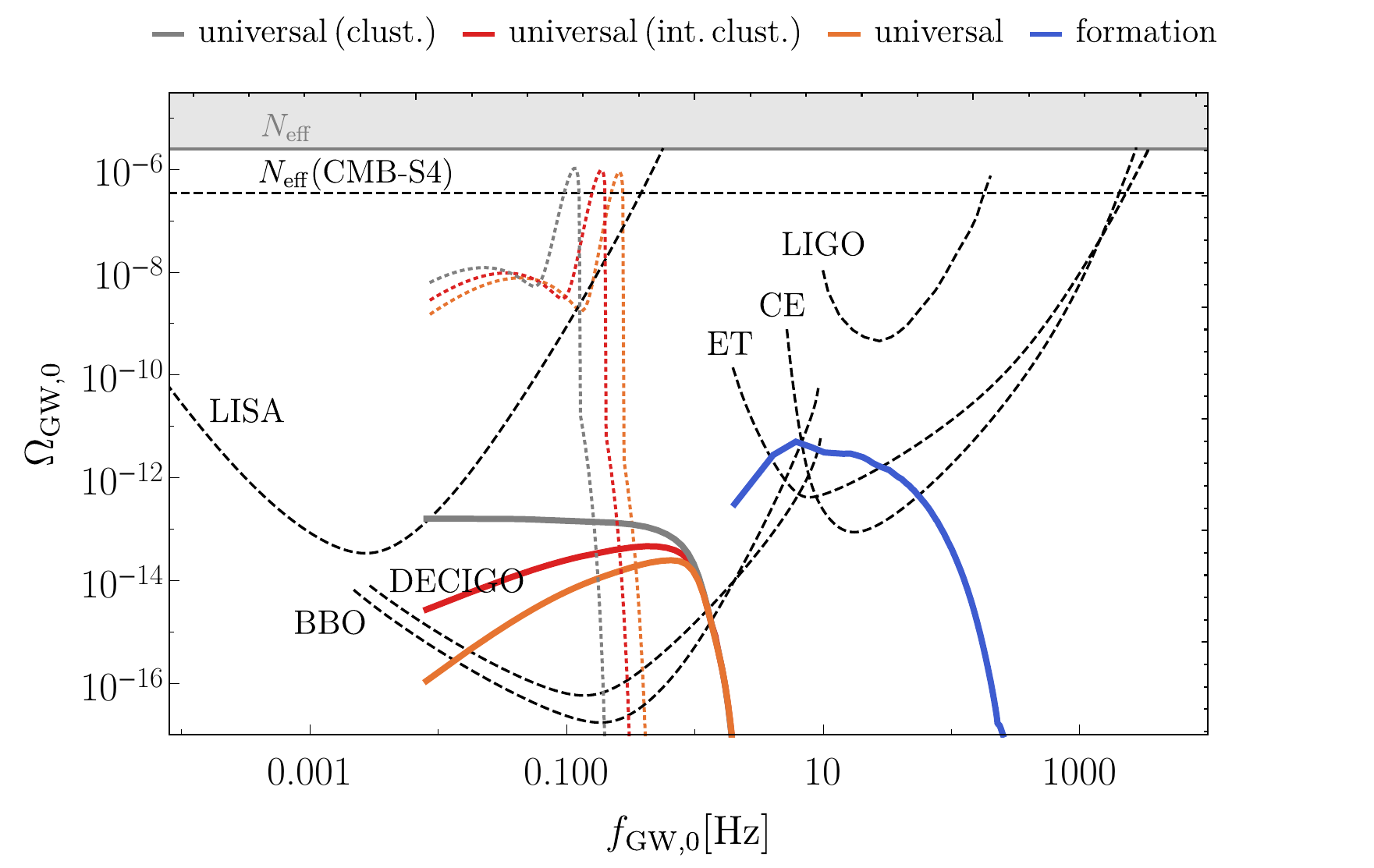}
\caption{\textbf{[Left]} The oscillon isocurvature power spectrum at the time
of formation (solid) and at the time of decay (dashed black). Power spectra for the cases of gravitational clustering (orange dash), intermediate clustering (orange and non-clustering (orange-black dashed) are also displayed. In the three cases the spectrum associated with the non-linear scales, $k/k_{nl}>1$ that correspond to the fixed
physical size of the oscillon objects, is the same. For more details on its evolution see \cite{Lozanov:2023aez}. \textbf{[Right]} The GW background from oscillon formation \cite{Lozanov:2019ylm} (blue), and the novel UGWs due to oscillon isocurvature (solid red) for the model parameters $k_{nl}\tau_i=10$, $k_{osc}\tau_i=10^2$, $k_{nl}=10^2k_{eq}$, and $m=10^{-18}m_{pl}$ induce additional GWs at second order~\cite{Lozanov:2022yoy}. Induced GWs associated with 
rapid transition from an early ALP oscillon matter domination to radiation domination phases due to oscillon decays~\cite{Lozanov:2022yoy} are shown by dotted lines. The estimated induced GW signals are from the linear part of $\mathcal{P}_S$, with $0<k<k_l$, for UV cut-offs, $\mathcal{P}_S(k_l)=10^{-2}$. We also display existing constraints from CMB and BBN originating from relativistic degrees of freedom $N_{\rm eff}$~\cite{Planck:2018vyg}, as well as sensitivity of CMB-S4~\cite{Abazajian:2019eic}, LISA, Einstein Telescope (ET),
Cosmic Explorer (CE), Big Bang Observer (BBO) and DECIGO~\cite{Schmitz:2020syl}, and LIGO O5~\cite{LIGOScientific:2016fpe}.
}
    \label{fig:GWPSEvolve}
\end{figure*}

The emergence of this expected $r$ power dependence can be heuristically seen as follows. First, consider virial theorem in FLRW Universe that is relevant for analysis of galaxies~\cite{1963ApJ...138..174L}
\Beq
\frac{d}{dt}(T+U)+H(2T+U)=0~.
\Eeq
Note when $T+U={\rm const}.$, the cloud of ``particles'', is virialised with $T=-U/2$. 

Consider now a Universe filled with soliton gas. On scales smaller than $r_1$, solitons are taken to be virialised and self-gravitating. On scales smaller than $r_0$ (with $r_0\ll r_1$) the point-particle approximation for solitons breaks down. The energy of a virialised cloud of solitons is thus
\Beq
E=T+U=\frac{U}{2}=-\frac{GM^2}{4}\int_{r_0}^{r_1}\frac{\xi(r)}{r}4\pi r^2dr
\Eeq
Assume $\xi(r)=A r^{x-2}$. Hence,
\Beq
E(x)=-\frac{\pi GM^2 A}{x} (r_1^x-r_0^x)~.
\Eeq
For $|x|\ll1$ there are 4 extrema for $E(x)$. Two extrema at $x=0$ exactly, which are unphysical, and two which are finite and tend to (see the case of galaxy clustering \cite{1980ApJ...235..299S})
\Beq
\begin{cases}
& \lim_{r_0\rightarrow 0} x_I = \dfrac{1}{\ln r_1} \\
&\lim_{r_0\rightarrow 0} x_{II} = \dfrac{2}{\ln r_0}\rightarrow 0 \\
\end{cases}
\Eeq
Here, $x_I$ is a maximum, $x_{II}$ is a minimum of $E(x)$. Hence, the inverse square power-law $\xi\propto r^{x_{II}-2}\rightarrow r^{-2}$, minimizes the energy of the virialised solitons.

The modification of large scale correlations due to interactions alters the power spectrum of solitons compared to the non-interacting case. While accurate modeling of the time evolution of the power spectrum is non-trivial and often requires detailed simulations, we can account for its asymptotic behavior at earlier times (i.e. around formation) and later times (i.e.~after several e-folds for the slow gravitational interactions to become efficient). In particular, on large scales, $k\ll k_{nl}$, around soliton formation, $\xi(r)\propto \delta(r)$, $\mathcal{P}_S\propto k^3$, and after soliton gravitational clustering, $\xi(r)\propto r^{-2}$, $\mathcal{P}_S\propto k^2$.

To estimate the degree to which the UGW signal is affected by a change in the large-scale power spectrum due to soliton interactions, we calculate UGWs in the several distinct scenarios as displayed in Fig.~\ref{fig:GWPSEvolve} (left panel). First, we consider UGWs when they are sourced by the spectrum associated with gravitational clustering with $\mathcal{P}_S=(k/k_{nl})^2$, the spectrum associated with
unclustered regime $\mathcal{P}_S=(k/k_{nl})^{3}$, as well as the spectrum associated with an
intermediate scenario with $\mathcal{P}_S=(k/k_{nl})^{5/2}$. The intermediate case has power that lies between the unclustered and gravitationally clustered regimes. As we will demonstrate, these spectra lead to significant modifications of the large-scale structure associated with the solitons and can dramatically enhance the UGW signal. We note that our results can be generalized to account for possible time dependence of the power spectra, which depends on model details, by considering interpolations between different limiting regimes.

As already noted, the gravitational clustering of solitons corresponds to nonlinear effects. On nonlinear scales, comparable to the soliton separation, $k\geq k_{nl}$, the inverse square correlation, which implies $\mathcal{P}_S\propto k^2$, can be understood in terms of virialisation of gravitationally bound soliton clusters. Furthermore, on quasi-linear scales, $k\ll k_{nl}$, relevant to the generation of UGWs, second-order gravitational interactions in the soliton fluid near the soliton domination stage drive the power spectrum of the soliton fluid perturbations to $\mathcal{P}_S\propto k^2$ \cite{Makino:1991rp}. 

Our discussion has broader implications beyond solitons. The clustering of solitons is analogous to clustering associated with a gas of primordial black holes that interact through gravity and can be formed throughout the early Universe~(e.g.~\cite{Suyama:2019cst,Domenech:2020ssp}). Further, our analysis can be generalized to scenarios with other long-range interactions beyond gravity. One such example is interaction mediated by a fifth Yukawa-type force. 
Since Yukawa long-range force (i.e. a force mediated by a nearly-massless particle) also scales as gravitational force $\propto r^{-2}$, this can be accommodated in our discussion by modifying gravitational coupling as $G \rightarrow (1 + \beta) G$ where $\beta$ is associated with the strength of Yukawa force (for recent discussion of structure formation and cosmology with Yukawa-interacting particles see~\cite{Domenech:2021uyx,Domenech:2023afs}).
 
\textit{Interacting ALP oscillons.}--- 
We illustrate the effect of soliton interactions on universal GW signatures in the context of axion-like particle (ALP) oscillons formed from an initially misaligned ALP field~(e.g.~\cite{Hui:2016ltb,Cyncynates:2022wlq}).
We consider that oscillon pseudo-solitons are formed from fragmentation of an ALP field $\phi$, described by canonical action
\begin{align}
S_\phi=&~\int d^4x\sqrt{-g}\left[\frac{1}{2}(\partial\phi)^2-V(\phi)\right]\,,\\
V(\phi)=&~m^2F^2\left[1-\cos\left(\frac{\phi}{F}\right)\right]\,,
\end{align}
here $V(\phi)$ is the ALP potential, $m$ is its mass and $F$ is the axion decay constant.
During inflation $m\ll H_{\rm inf}$, where $H_{\rm inf}$ is the Hubble parameter, and the light field $\phi$ is misaligned from origin as $\phi_{\rm inf} \sim F$. Subsequently, after inflation $\phi$ begins to oscillate when $m$ overcomes the Hubble rate, when the Universe is radiation dominated and $\phi$ is subdominant.
Oscillating $\phi$ rapidly fragments into solitonic oscillon lumps due to instabilities (e.g.~\cite{Amin:2011hj}).

The energy density power spectrum of $\phi$ contains several distinct peaks. The first peak at $k_{osc}(\tau)\sim a(\tau)m$ is associated with a physical wavenumber   corresponding to oscillon size $\sim m^{-1}$. The second peak is located at $k_{nl}\sim10^{-1}m/a_i={\rm const.}\,$, corresponding to co-moving separation between oscillons.  

For ALP oscillons that are approximately non-interacting~\cite{Lozanov:2023aez}, for $k < k_{nl}$ the $\phi$ energy density power
spectrum follows Poissonian distribution $\propto k^3$. On very long super-horizon scales $k \ll k_{\rm nl}$ nearly scale invariant perturbations are constrained by CMB bounds on isocurvature~\cite{Planck:2018vyg}, which are saturated if $F \sim 10^5 H_{inf}$~\cite{Lozanov:2023aez}. Starting with an ALP field that is spectator during inflation and that starts oscillating when $H \sim m$ and fragments into oscillons around $H_i \lesssim 10^{-2} m$~\cite{Amin:2011hj,Fukunaga:2019unq}, one can show that ~\cite{Lozanov:2023aez}
\Beq \label{eq:malp}
m^2\sim m_{pl}^2\frac{k_{\rm eq}}{\mathcal{H}_i}\frac{H_i^2}{F^2}\,~,
\Eeq
where $m_{pl}$ is the Planck scale and $\mathcal{H} = H a$. 

In Fig.~\ref{fig:GWPSEvolve} we display (right panel) the results for UGWs, calculated following procedure of Ref.~\cite{Lozanov:2023aez} and as summarized above, considering the power spectra shown on Fig.~1 (left panel). We also display UGWs for the case of ``no clustering'', gravitationally ``clustered'' solitons, as well as an ``intermediate clustering'' case, with power spectrum lying between them. 
Our findings illustrate that the resulting lower frequency tail of the UGW background is enhanced for power spectra corresponding to increased soliton clustering, due to the increased correlations in the soliton density field such as that originating from gravitational interactions between the objects. We expect that the flat UGW spectrum for the case of clustered solitons will have an IR cut-off at the horizon scale at the time of decay. Below the IR cut-off the GW power spectrum is expected to follow $\propto f^3$. Properly capturing this low-frequency behaviour in detail requires complex computations beyond the scope of this work, and we leave it for future investigation. Our calculations capture the dominant signal contributions, which can be probed by LISA or other expeirments for some of the parameters chosen in Fig.~\ref{fig:GWPSEvolve}. 

In addition to UGWs signals, in Fig.~\ref{fig:GWPSEvolve} we display GWs expected to be generated during oscillon formation~\cite{Lozanov:2019ylm}, with characteristic peak frequency being approximately at $k_{osc}$ and power of
\Beq
\Omega_{\rm GW}(f\sim k_{osc}(\tau_i))\sim (\delta^{TT})^2\frac{\mathcal{H}_i^2}{k_{osc}^2(\tau_i)}\,,
\Eeq
where the dimensionless energy momentum tensor is $\delta^{TT} \lesssim \mathcal{O}(1)\times (k_{\rm eq}\tau_i)$. Further, we display in Fig.~\ref{fig:GWPSEvolve} (dashed) induced GWs originating from oscillon decays~\cite{Lozanov:2022yoy}, which can be significant depending on the model and regimes considered. Here, we considered signal from linear part of $\mathcal{P}_S$, with $0 < k < k_l$, for ultraviolet (UV) cut-offs, $\mathcal{P}_S (k_l) = 10^{-2}$.

Correlations of GW signatures in the lower frequency bands with higher frequency bands allow to distinguish different scenarios and models of cosmological solitons and their interactions.

{\it Conclusions}.--- Formation of solitonic objects in the early Universe is expected to be generically accompanied by background of UGWs with frequencies smaller than scales of nonlinearities, stemming from isocurvature perturbations due to soliton population distribution. As we demonstrated, soliton interactions can modify their distribution and associated power spectrum. A natural realization of this is soliton 
clustering due to gravitational interactions.
We found that interactions among solitons, and in particular gravitational clustering, can significantly enhance produced UGWs. For the case of ALP oscillons, the enhancement of UGWs due to soliton gravitational clustering effect can be orders of magnitude in the frequencies relevant for upcoming GW experiments. 
We identify that correlated GW signals across multiple frequency bands will allow to distinguish theories and different soliton interactions.
Our results are 
applicable for broad classes of cosmological scenarios with soliton formation, including those where solitons are interacting through a long-range Yukawa-type fifth force.

~\newline
{\it Acknowledgments}.
This work is supported by World Premier International Research Center Initiative (WPI), MEXT, Japan. This work is also
supported in part by the JSPS KAKENHI grant Nos. 20H05853 (M.S.)
and 23K13109 (V.T.).

\bibliography{IGWSIsoOscillons}

\begin{thebibliography}{30}%
\makeatletter
\providecommand \@ifxundefined [1]{%
 \@ifx{#1\undefined}
}%
\providecommand \@ifnum [1]{%
 \ifnum #1\expandafter \@firstoftwo
 \else \expandafter \@secondoftwo
 \fi
}%
\providecommand \@ifx [1]{%
 \ifx #1\expandafter \@firstoftwo
 \else \expandafter \@secondoftwo
 \fi
}%
\providecommand \natexlab [1]{#1}%
\providecommand \enquote  [1]{``#1''}%
\providecommand \bibnamefont  [1]{#1}%
\providecommand \bibfnamefont [1]{#1}%
\providecommand \citenamefont [1]{#1}%
\providecommand \href@noop [0]{\@secondoftwo}%
\providecommand \href [0]{\begingroup \@sanitize@url \@href}%
\providecommand \@href[1]{\@@startlink{#1}\@@href}%
\providecommand \@@href[1]{\endgroup#1\@@endlink}%
\providecommand \@sanitize@url [0]{\catcode `\\12\catcode `\$12\catcode
  `\&12\catcode `\#12\catcode `\^12\catcode `\_12\catcode `\%12\relax}%
\providecommand \@@startlink[1]{}%
\providecommand \@@endlink[0]{}%
\providecommand \url  [0]{\begingroup\@sanitize@url \@url }%
\providecommand \@url [1]{\endgroup\@href {#1}{\urlprefix }}%
\providecommand \urlprefix  [0]{URL }%
\providecommand \Eprint [0]{\href }%
\providecommand \doibase [0]{http://dx.doi.org/}%
\providecommand \selectlanguage [0]{\@gobble}%
\providecommand \bibinfo  [0]{\@secondoftwo}%
\providecommand \bibfield  [0]{\@secondoftwo}%
\providecommand \translation [1]{[#1]}%
\providecommand \BibitemOpen [0]{}%
\providecommand \bibitemStop [0]{}%
\providecommand \bibitemNoStop [0]{.\EOS\space}%
\providecommand \EOS [0]{\spacefactor3000\relax}%
\providecommand \BibitemShut  [1]{\csname bibitem#1\endcsname}%
\let\auto@bib@innerbib\@empty
\bibitem [{\citenamefont {Vilenkin}\ and\ \citenamefont
  {Shellard}(2000)}]{Vilenkin:2000jqa}%
  \BibitemOpen
  \bibfield  {author} {\bibinfo {author} {\bibfnamefont {A.}~\bibnamefont
  {Vilenkin}}\ and\ \bibinfo {author} {\bibfnamefont {E.~P.~S.}\ \bibnamefont
  {Shellard}},\ }\href@noop {} {\emph {\bibinfo {title} {{Cosmic Strings and
  Other Topological Defects}}}}\ (\bibinfo  {publisher} {Cambridge University
  Press},\ \bibinfo {year} {2000})\BibitemShut {NoStop}%
\bibitem [{\citenamefont {Shnir}(2018)}]{Shnir:2018yzp}%
  \BibitemOpen
  \bibfield  {author} {\bibinfo {author} {\bibfnamefont {Y.~M.}\ \bibnamefont
  {Shnir}},\ }\href@noop {} {\emph {\bibinfo {title} {{Topological and
  Non-Topological Solitons in Scalar Field Theories}}}}\ (\bibinfo  {publisher}
  {Cambridge University Press},\ \bibinfo {year} {2018})\BibitemShut {NoStop}%
\bibitem [{\citenamefont {Kibble}(1976)}]{Kibble:1976sj}%
  \BibitemOpen
  \bibfield  {author} {\bibinfo {author} {\bibfnamefont {T.~W.~B.}\
  \bibnamefont {Kibble}},\ }\href {\doibase 10.1088/0305-4470/9/8/029}
  {\bibfield  {journal} {\bibinfo  {journal} {J. Phys. A}\ }\textbf {\bibinfo
  {volume} {9}},\ \bibinfo {pages} {1387} (\bibinfo {year} {1976})}\BibitemShut
  {NoStop}%
\bibitem [{\citenamefont {Zurek}(1985)}]{Zurek:1985qw}%
  \BibitemOpen
  \bibfield  {author} {\bibinfo {author} {\bibfnamefont {W.~H.}\ \bibnamefont
  {Zurek}},\ }\href {\doibase 10.1038/317505a0} {\bibfield  {journal} {\bibinfo
   {journal} {Nature}\ }\textbf {\bibinfo {volume} {317}},\ \bibinfo {pages}
  {505} (\bibinfo {year} {1985})}\BibitemShut {NoStop}%
\bibitem [{\citenamefont {Amin}\ \emph {et~al.}(2014)\citenamefont {Amin},
  \citenamefont {Hertzberg}, \citenamefont {Kaiser},\ and\ \citenamefont
  {Karouby}}]{Amin:2014eta}%
  \BibitemOpen
  \bibfield  {author} {\bibinfo {author} {\bibfnamefont {M.~A.}\ \bibnamefont
  {Amin}}, \bibinfo {author} {\bibfnamefont {M.~P.}\ \bibnamefont {Hertzberg}},
  \bibinfo {author} {\bibfnamefont {D.~I.}\ \bibnamefont {Kaiser}}, \ and\
  \bibinfo {author} {\bibfnamefont {J.}~\bibnamefont {Karouby}},\ }\href
  {\doibase 10.1142/S0218271815300037} {\bibfield  {journal} {\bibinfo
  {journal} {Int. J. Mod. Phys. D}\ }\textbf {\bibinfo {volume} {24}},\
  \bibinfo {pages} {1530003} (\bibinfo {year} {2014})},\ \Eprint
  {http://arxiv.org/abs/1410.3808} {arXiv:1410.3808 [hep-ph]} \BibitemShut
  {NoStop}%
\bibitem [{\citenamefont {Amin}\ \emph {et~al.}(2012)\citenamefont {Amin},
  \citenamefont {Easther}, \citenamefont {Finkel}, \citenamefont {Flauger},\
  and\ \citenamefont {Hertzberg}}]{Amin:2011hj}%
  \BibitemOpen
  \bibfield  {author} {\bibinfo {author} {\bibfnamefont {M.~A.}\ \bibnamefont
  {Amin}}, \bibinfo {author} {\bibfnamefont {R.}~\bibnamefont {Easther}},
  \bibinfo {author} {\bibfnamefont {H.}~\bibnamefont {Finkel}}, \bibinfo
  {author} {\bibfnamefont {R.}~\bibnamefont {Flauger}}, \ and\ \bibinfo
  {author} {\bibfnamefont {M.~P.}\ \bibnamefont {Hertzberg}},\ }\href {\doibase
  10.1103/PhysRevLett.108.241302} {\bibfield  {journal} {\bibinfo  {journal}
  {Phys. Rev. Lett.}\ }\textbf {\bibinfo {volume} {108}},\ \bibinfo {pages}
  {241302} (\bibinfo {year} {2012})},\ \Eprint {http://arxiv.org/abs/1106.3335}
  {arXiv:1106.3335 [astro-ph.CO]} \BibitemShut {NoStop}%
\bibitem [{\citenamefont {Cotner}\ \emph {et~al.}(2018)\citenamefont {Cotner},
  \citenamefont {Kusenko},\ and\ \citenamefont {Takhistov}}]{Cotner:2018vug}%
  \BibitemOpen
  \bibfield  {author} {\bibinfo {author} {\bibfnamefont {E.}~\bibnamefont
  {Cotner}}, \bibinfo {author} {\bibfnamefont {A.}~\bibnamefont {Kusenko}}, \
  and\ \bibinfo {author} {\bibfnamefont {V.}~\bibnamefont {Takhistov}},\ }\href
  {\doibase 10.1103/PhysRevD.98.083513} {\bibfield  {journal} {\bibinfo
  {journal} {Phys. Rev. D}\ }\textbf {\bibinfo {volume} {98}},\ \bibinfo
  {pages} {083513} (\bibinfo {year} {2018})},\ \Eprint
  {http://arxiv.org/abs/1801.03321} {arXiv:1801.03321 [astro-ph.CO]}
  \BibitemShut {NoStop}%
\bibitem [{\citenamefont {Cotner}\ \emph {et~al.}(2019)\citenamefont {Cotner},
  \citenamefont {Kusenko}, \citenamefont {Sasaki},\ and\ \citenamefont
  {Takhistov}}]{Cotner:2019ykd}%
  \BibitemOpen
  \bibfield  {author} {\bibinfo {author} {\bibfnamefont {E.}~\bibnamefont
  {Cotner}}, \bibinfo {author} {\bibfnamefont {A.}~\bibnamefont {Kusenko}},
  \bibinfo {author} {\bibfnamefont {M.}~\bibnamefont {Sasaki}}, \ and\ \bibinfo
  {author} {\bibfnamefont {V.}~\bibnamefont {Takhistov}},\ }\href {\doibase
  10.1088/1475-7516/2019/10/077} {\bibfield  {journal} {\bibinfo  {journal}
  {JCAP}\ }\textbf {\bibinfo {volume} {10}},\ \bibinfo {pages} {077} (\bibinfo
  {year} {2019})},\ \Eprint {http://arxiv.org/abs/1907.10613} {arXiv:1907.10613
  [astro-ph.CO]} \BibitemShut {NoStop}%
\bibitem [{\citenamefont {Lozanov}\ and\ \citenamefont
  {Amin}(2019)}]{Lozanov:2019ylm}%
  \BibitemOpen
  \bibfield  {author} {\bibinfo {author} {\bibfnamefont {K.~D.}\ \bibnamefont
  {Lozanov}}\ and\ \bibinfo {author} {\bibfnamefont {M.~A.}\ \bibnamefont
  {Amin}},\ }\href {\doibase 10.1103/PhysRevD.99.123504} {\bibfield  {journal}
  {\bibinfo  {journal} {Phys. Rev. D}\ }\textbf {\bibinfo {volume} {99}},\
  \bibinfo {pages} {123504} (\bibinfo {year} {2019})},\ \Eprint
  {http://arxiv.org/abs/1902.06736} {arXiv:1902.06736 [astro-ph.CO]}
  \BibitemShut {NoStop}%
\bibitem [{\citenamefont {Caprini}\ and\ \citenamefont
  {Figueroa}(2018)}]{Caprini:2018mtu}%
  \BibitemOpen
  \bibfield  {author} {\bibinfo {author} {\bibfnamefont {C.}~\bibnamefont
  {Caprini}}\ and\ \bibinfo {author} {\bibfnamefont {D.~G.}\ \bibnamefont
  {Figueroa}},\ }\href {\doibase 10.1088/1361-6382/aac608} {\bibfield
  {journal} {\bibinfo  {journal} {Class. Quant. Grav.}\ }\textbf {\bibinfo
  {volume} {35}},\ \bibinfo {pages} {163001} (\bibinfo {year} {2018})},\
  \Eprint {http://arxiv.org/abs/1801.04268} {arXiv:1801.04268 [astro-ph.CO]}
  \BibitemShut {NoStop}%
\bibitem [{\citenamefont {Lozanov}\ \emph {et~al.}(2023)\citenamefont
  {Lozanov}, \citenamefont {Sasaki},\ and\ \citenamefont
  {Takhistov}}]{Lozanov:2023aez}%
  \BibitemOpen
  \bibfield  {author} {\bibinfo {author} {\bibfnamefont {K.~D.}\ \bibnamefont
  {Lozanov}}, \bibinfo {author} {\bibfnamefont {M.}~\bibnamefont {Sasaki}}, \
  and\ \bibinfo {author} {\bibfnamefont {V.}~\bibnamefont {Takhistov}},\
  }\href@noop {} {\  (\bibinfo {year} {2023})},\ \Eprint
  {http://arxiv.org/abs/2304.06709} {arXiv:2304.06709 [astro-ph.CO]}
  \BibitemShut {NoStop}%
\bibitem [{\citenamefont {Kodama}\ and\ \citenamefont
  {Sasaki}(1984)}]{Kodama:1984ziu}%
  \BibitemOpen
  \bibfield  {author} {\bibinfo {author} {\bibfnamefont {H.}~\bibnamefont
  {Kodama}}\ and\ \bibinfo {author} {\bibfnamefont {M.}~\bibnamefont
  {Sasaki}},\ }\href {\doibase 10.1143/PTPS.78.1} {\bibfield  {journal}
  {\bibinfo  {journal} {Prog. Theor. Phys. Suppl.}\ }\textbf {\bibinfo {volume}
  {78}},\ \bibinfo {pages} {1} (\bibinfo {year} {1984})}\BibitemShut {NoStop}%
\bibitem [{\citenamefont {Kodama}\ and\ \citenamefont
  {Sasaki}(1987)}]{Kodama:1986ud}%
  \BibitemOpen
  \bibfield  {author} {\bibinfo {author} {\bibfnamefont {H.}~\bibnamefont
  {Kodama}}\ and\ \bibinfo {author} {\bibfnamefont {M.}~\bibnamefont
  {Sasaki}},\ }\href {\doibase 10.1142/S0217751X8700020X} {\bibfield  {journal}
  {\bibinfo  {journal} {Int. J. Mod. Phys. A}\ }\textbf {\bibinfo {volume}
  {2}},\ \bibinfo {pages} {491} (\bibinfo {year} {1987})}\BibitemShut {NoStop}%
\bibitem [{\citenamefont {Dom\`enech}\ \emph {et~al.}(2022)\citenamefont
  {Dom\`enech}, \citenamefont {Passaglia},\ and\ \citenamefont
  {Renaux-Petel}}]{Domenech:2021and}%
  \BibitemOpen
  \bibfield  {author} {\bibinfo {author} {\bibfnamefont {G.}~\bibnamefont
  {Dom\`enech}}, \bibinfo {author} {\bibfnamefont {S.}~\bibnamefont
  {Passaglia}}, \ and\ \bibinfo {author} {\bibfnamefont {S.}~\bibnamefont
  {Renaux-Petel}},\ }\href {\doibase 10.1088/1475-7516/2022/03/023} {\bibfield
  {journal} {\bibinfo  {journal} {JCAP}\ }\textbf {\bibinfo {volume} {03}},\
  \bibinfo {pages} {023} (\bibinfo {year} {2022})},\ \Eprint
  {http://arxiv.org/abs/2112.10163} {arXiv:2112.10163 [astro-ph.CO]}
  \BibitemShut {NoStop}%
\bibitem [{\citenamefont {Aghanim}\ \emph {et~al.}(2020)\citenamefont {Aghanim}
  \emph {et~al.}}]{Planck:2018vyg}%
  \BibitemOpen
  \bibfield  {author} {\bibinfo {author} {\bibfnamefont {N.}~\bibnamefont
  {Aghanim}} \emph {et~al.} (\bibinfo {collaboration} {Planck}),\ }\href
  {\doibase 10.1051/0004-6361/201833910} {\bibfield  {journal} {\bibinfo
  {journal} {Astron. Astrophys.}\ }\textbf {\bibinfo {volume} {641}},\ \bibinfo
  {pages} {A6} (\bibinfo {year} {2020})},\ \bibinfo {note} {[Erratum:
  Astron.Astrophys. 652, C4 (2021)]},\ \Eprint
  {http://arxiv.org/abs/1807.06209} {arXiv:1807.06209 [astro-ph.CO]}
  \BibitemShut {NoStop}%
\bibitem [{\citenamefont {Amin}\ and\ \citenamefont
  {Mocz}(2019)}]{Amin:2019ums}%
  \BibitemOpen
  \bibfield  {author} {\bibinfo {author} {\bibfnamefont {M.~A.}\ \bibnamefont
  {Amin}}\ and\ \bibinfo {author} {\bibfnamefont {P.}~\bibnamefont {Mocz}},\
  }\href {\doibase 10.1103/PhysRevD.100.063507} {\bibfield  {journal} {\bibinfo
   {journal} {Phys. Rev. D}\ }\textbf {\bibinfo {volume} {100}},\ \bibinfo
  {pages} {063507} (\bibinfo {year} {2019})},\ \Eprint
  {http://arxiv.org/abs/1902.07261} {arXiv:1902.07261 [astro-ph.CO]}
  \BibitemShut {NoStop}%
\bibitem [{\citenamefont {Lozanov}\ and\ \citenamefont
  {Takhistov}(2023)}]{Lozanov:2022yoy}%
  \BibitemOpen
  \bibfield  {author} {\bibinfo {author} {\bibfnamefont {K.~D.}\ \bibnamefont
  {Lozanov}}\ and\ \bibinfo {author} {\bibfnamefont {V.}~\bibnamefont
  {Takhistov}},\ }\href {\doibase 10.1103/PhysRevLett.130.181002} {\bibfield
  {journal} {\bibinfo  {journal} {Phys. Rev. Lett.}\ }\textbf {\bibinfo
  {volume} {130}},\ \bibinfo {pages} {181002} (\bibinfo {year} {2023})},\
  \Eprint {http://arxiv.org/abs/2204.07152} {arXiv:2204.07152 [astro-ph.CO]}
  \BibitemShut {NoStop}%
\bibitem [{\citenamefont {Abazajian}\ \emph {et~al.}(2019)\citenamefont
  {Abazajian} \emph {et~al.}}]{Abazajian:2019eic}%
  \BibitemOpen
  \bibfield  {author} {\bibinfo {author} {\bibfnamefont {K.}~\bibnamefont
  {Abazajian}} \emph {et~al.},\ }\href@noop {} {\  (\bibinfo {year} {2019})},\
  \Eprint {http://arxiv.org/abs/1907.04473} {arXiv:1907.04473 [astro-ph.IM]}
  \BibitemShut {NoStop}%
\bibitem [{\citenamefont {Schmitz}(2021)}]{Schmitz:2020syl}%
  \BibitemOpen
  \bibfield  {author} {\bibinfo {author} {\bibfnamefont {K.}~\bibnamefont
  {Schmitz}},\ }\href {\doibase 10.1007/JHEP01(2021)097} {\bibfield  {journal}
  {\bibinfo  {journal} {JHEP}\ }\textbf {\bibinfo {volume} {01}},\ \bibinfo
  {pages} {097} (\bibinfo {year} {2021})},\ \Eprint
  {http://arxiv.org/abs/2002.04615} {arXiv:2002.04615 [hep-ph]} \BibitemShut
  {NoStop}%
\bibitem [{\citenamefont {Abbott}\ \emph {et~al.}(2016)\citenamefont {Abbott}
  \emph {et~al.}}]{LIGOScientific:2016fpe}%
  \BibitemOpen
  \bibfield  {author} {\bibinfo {author} {\bibfnamefont {B.~P.}\ \bibnamefont
  {Abbott}} \emph {et~al.} (\bibinfo {collaboration} {LIGO Scientific,
  Virgo}),\ }\href {\doibase 10.1103/PhysRevLett.116.131102} {\bibfield
  {journal} {\bibinfo  {journal} {Phys. Rev. Lett.}\ }\textbf {\bibinfo
  {volume} {116}},\ \bibinfo {pages} {131102} (\bibinfo {year} {2016})},\
  \Eprint {http://arxiv.org/abs/1602.03847} {arXiv:1602.03847 [gr-qc]}
  \BibitemShut {NoStop}%
\bibitem [{\citenamefont {{Layzer}}(1963)}]{1963ApJ...138..174L}%
  \BibitemOpen
  \bibfield  {author} {\bibinfo {author} {\bibfnamefont {D.}~\bibnamefont
  {{Layzer}}},\ }\href {\doibase 10.1086/147625} {\bibfield  {journal}
  {\bibinfo  {journal} {\apj}\ }\textbf {\bibinfo {volume} {138}},\ \bibinfo
  {pages} {174} (\bibinfo {year} {1963})}\BibitemShut {NoStop}%
\bibitem [{\citenamefont {{Saslaw}}(1980)}]{1980ApJ...235..299S}%
  \BibitemOpen
  \bibfield  {author} {\bibinfo {author} {\bibfnamefont {W.~C.}\ \bibnamefont
  {{Saslaw}}},\ }\href {\doibase 10.1086/157634} {\bibfield  {journal}
  {\bibinfo  {journal} {\apj}\ }\textbf {\bibinfo {volume} {235}},\ \bibinfo
  {pages} {299} (\bibinfo {year} {1980})}\BibitemShut {NoStop}%
\bibitem [{\citenamefont {Makino}\ \emph {et~al.}(1992)\citenamefont {Makino},
  \citenamefont {Sasaki},\ and\ \citenamefont {Suto}}]{Makino:1991rp}%
  \BibitemOpen
  \bibfield  {author} {\bibinfo {author} {\bibfnamefont {N.}~\bibnamefont
  {Makino}}, \bibinfo {author} {\bibfnamefont {M.}~\bibnamefont {Sasaki}}, \
  and\ \bibinfo {author} {\bibfnamefont {Y.}~\bibnamefont {Suto}},\ }\href
  {\doibase 10.1103/PhysRevD.46.585} {\bibfield  {journal} {\bibinfo  {journal}
  {Phys. Rev. D}\ }\textbf {\bibinfo {volume} {46}},\ \bibinfo {pages} {585}
  (\bibinfo {year} {1992})}\BibitemShut {NoStop}%
\bibitem [{\citenamefont {Suyama}\ and\ \citenamefont
  {Yokoyama}(2019)}]{Suyama:2019cst}%
  \BibitemOpen
  \bibfield  {author} {\bibinfo {author} {\bibfnamefont {T.}~\bibnamefont
  {Suyama}}\ and\ \bibinfo {author} {\bibfnamefont {S.}~\bibnamefont
  {Yokoyama}},\ }\href {\doibase 10.1093/ptep/ptz105} {\bibfield  {journal}
  {\bibinfo  {journal} {PTEP}\ }\textbf {\bibinfo {volume} {2019}},\ \bibinfo
  {pages} {103E02} (\bibinfo {year} {2019})},\ \Eprint
  {http://arxiv.org/abs/1906.04958} {arXiv:1906.04958 [astro-ph.CO]}
  \BibitemShut {NoStop}%
\bibitem [{\citenamefont {Domenech}\ \emph {et~al.}(2021)\citenamefont
  {Domenech}, \citenamefont {Lin},\ and\ \citenamefont
  {Sasaki}}]{Domenech:2020ssp}%
  \BibitemOpen
  \bibfield  {author} {\bibinfo {author} {\bibfnamefont {G.}~\bibnamefont
  {Domenech}}, \bibinfo {author} {\bibfnamefont {C.}~\bibnamefont {Lin}}, \
  and\ \bibinfo {author} {\bibfnamefont {M.}~\bibnamefont {Sasaki}},\ }\href
  {\doibase 10.1088/1475-7516/2021/11/E01} {\bibfield  {journal} {\bibinfo
  {journal} {JCAP}\ }\textbf {\bibinfo {volume} {04}},\ \bibinfo {pages} {062}
  (\bibinfo {year} {2021})},\ \bibinfo {note} {[Erratum: JCAP 11, E01
  (2021)]},\ \Eprint {http://arxiv.org/abs/2012.08151} {arXiv:2012.08151
  [gr-qc]} \BibitemShut {NoStop}%
\bibitem [{\citenamefont {Dom\`enech}\ and\ \citenamefont
  {Sasaki}(2021)}]{Domenech:2021uyx}%
  \BibitemOpen
  \bibfield  {author} {\bibinfo {author} {\bibfnamefont {G.}~\bibnamefont
  {Dom\`enech}}\ and\ \bibinfo {author} {\bibfnamefont {M.}~\bibnamefont
  {Sasaki}},\ }\href {\doibase 10.1088/1475-7516/2021/06/030} {\bibfield
  {journal} {\bibinfo  {journal} {JCAP}\ }\textbf {\bibinfo {volume} {06}},\
  \bibinfo {pages} {030} (\bibinfo {year} {2021})},\ \Eprint
  {http://arxiv.org/abs/2104.05271} {arXiv:2104.05271 [hep-th]} \BibitemShut
  {NoStop}%
\bibitem [{\citenamefont {Dom\`enech}\ \emph {et~al.}(2023)\citenamefont
  {Dom\`enech}, \citenamefont {Inman}, \citenamefont {Kusenko},\ and\
  \citenamefont {Sasaki}}]{Domenech:2023afs}%
  \BibitemOpen
  \bibfield  {author} {\bibinfo {author} {\bibfnamefont {G.}~\bibnamefont
  {Dom\`enech}}, \bibinfo {author} {\bibfnamefont {D.}~\bibnamefont {Inman}},
  \bibinfo {author} {\bibfnamefont {A.}~\bibnamefont {Kusenko}}, \ and\
  \bibinfo {author} {\bibfnamefont {M.}~\bibnamefont {Sasaki}},\ }\href@noop {}
  {\  (\bibinfo {year} {2023})},\ \Eprint {http://arxiv.org/abs/2304.13053}
  {arXiv:2304.13053 [astro-ph.CO]} \BibitemShut {NoStop}%
\bibitem [{\citenamefont {Hui}\ \emph {et~al.}(2017)\citenamefont {Hui},
  \citenamefont {Ostriker}, \citenamefont {Tremaine},\ and\ \citenamefont
  {Witten}}]{Hui:2016ltb}%
  \BibitemOpen
  \bibfield  {author} {\bibinfo {author} {\bibfnamefont {L.}~\bibnamefont
  {Hui}}, \bibinfo {author} {\bibfnamefont {J.~P.}\ \bibnamefont {Ostriker}},
  \bibinfo {author} {\bibfnamefont {S.}~\bibnamefont {Tremaine}}, \ and\
  \bibinfo {author} {\bibfnamefont {E.}~\bibnamefont {Witten}},\ }\href
  {\doibase 10.1103/PhysRevD.95.043541} {\bibfield  {journal} {\bibinfo
  {journal} {Phys. Rev. D}\ }\textbf {\bibinfo {volume} {95}},\ \bibinfo
  {pages} {043541} (\bibinfo {year} {2017})},\ \Eprint
  {http://arxiv.org/abs/1610.08297} {arXiv:1610.08297 [astro-ph.CO]}
  \BibitemShut {NoStop}%
\bibitem [{\citenamefont {Cyncynates}\ \emph {et~al.}(2022)\citenamefont
  {Cyncynates}, \citenamefont {Simon}, \citenamefont {Thompson},\ and\
  \citenamefont {Weiner}}]{Cyncynates:2022wlq}%
  \BibitemOpen
  \bibfield  {author} {\bibinfo {author} {\bibfnamefont {D.}~\bibnamefont
  {Cyncynates}}, \bibinfo {author} {\bibfnamefont {O.}~\bibnamefont {Simon}},
  \bibinfo {author} {\bibfnamefont {J.~O.}\ \bibnamefont {Thompson}}, \ and\
  \bibinfo {author} {\bibfnamefont {Z.~J.}\ \bibnamefont {Weiner}},\ }\href
  {\doibase 10.1103/PhysRevD.106.083503} {\bibfield  {journal} {\bibinfo
  {journal} {Phys. Rev. D}\ }\textbf {\bibinfo {volume} {106}},\ \bibinfo
  {pages} {083503} (\bibinfo {year} {2022})},\ \Eprint
  {http://arxiv.org/abs/2208.05501} {arXiv:2208.05501 [hep-ph]} \BibitemShut
  {NoStop}%
\bibitem [{\citenamefont {Fukunaga}\ \emph {et~al.}(2019)\citenamefont
  {Fukunaga}, \citenamefont {Kitajima},\ and\ \citenamefont
  {Urakawa}}]{Fukunaga:2019unq}%
  \BibitemOpen
  \bibfield  {author} {\bibinfo {author} {\bibfnamefont {H.}~\bibnamefont
  {Fukunaga}}, \bibinfo {author} {\bibfnamefont {N.}~\bibnamefont {Kitajima}},
  \ and\ \bibinfo {author} {\bibfnamefont {Y.}~\bibnamefont {Urakawa}},\ }\href
  {\doibase 10.1088/1475-7516/2019/06/055} {\bibfield  {journal} {\bibinfo
  {journal} {JCAP}\ }\textbf {\bibinfo {volume} {06}},\ \bibinfo {pages} {055}
  (\bibinfo {year} {2019})},\ \Eprint {http://arxiv.org/abs/1903.02119}
  {arXiv:1903.02119 [astro-ph.CO]} \BibitemShut {NoStop}%
\end{thebibliography}%

\end{document}